# Crowdsourcing for Participatory Democracies: Efficient Elicitation of Social Choice Functions

DAVID T. LEE, ASHISH GOEL, Stanford University
TANJA AITAMURTO, UC Berkeley and University of Tampere
HELENE LANDEMORE, Yale University

Recent years have seen an increase in democratic innovations[Smith 2009] aimed at increasing the participation of the public in policy-making. This observation, coupled with the increasing prevalence of internet-based communication, points to a very real possibility of implementing participatory democracies on a mass-scale in which every individual is invited to contribute their ideas and opinions[Salganik and Levy 2012].

One important question in implementing crowdsourcing experiments of this type is the aggregation problem: given a large number of ideas, how can one identify the top ideas without requiring any individual, whether an appointed government expert or a participant, to spend too much time or effort in the evaluation process? A natural approach to aggregation in the democratic setting is to use voting rules, also known as social choice functions[Brandt et al. 2012], to find the top ideas.

In this paper, we present theoretical and empirical results indicating the usefulness of voting rules for participatory democracies. We first give algorithms which efficiently elicit $\epsilon$-approximations to two prominent voting rules: the Borda rule and the Condorcet winner. This result circumvents previous prohibitive lower bounds[Conitzer and Sandholm 2005][Service and Adams 2012] and is surprisingly strong: even if the number of ideas is as large as the number of participants, each participant will only have to make a logarithmic number of comparisons, an exponential improvement over the linear number of comparisons previously needed. Essentially, we show that these voting rules, which scale inefficiently, can be easy to implement when the winner wins by a margin or an approximation suffices.

We demonstrate the approach in an experiment in Finland's recent off-road traffic law reform[Aitamurto and Landemore 2013]. The Finnish experiment, as we will refer to it from this point on, engaged the Finnish people in 1) identifying problems, 2) proposing solutions, and 3) evaluating ideas for the off-road traffic law. In the evaluation stage, 308 participants took part in ranking ideas in 41 different topics, each of which had a number of ideas ranging from 2 to 15. For $\epsilon = 0.05$ and $0.1$, we show that the total number of comparisons needed for all participants is linear in the number of ideas and that the constant is not large. As an example, when the derived linear trend is extrapolated to the case of aggregating 100 ideas with 1000 participants, an error of $0.05$ only requires each participant to make $19$ comparisons. If we only need an error of $0.1$, then $8$ comparisons suffice.

Finally, we note a few other experimental observations which support the use of voting rules for aggregation. First, we observe that rating, one of the common alternatives to ranking, manifested effects of bias in our data. Second, we show that very few of the topics lacked a Condorcet winner, one of the prominent negative results in voting. Finally, we show data hinting at a potential future direction: the use of partial rankings as opposed to pairwise comparisons to further decrease the elicitation time.

## 1. A SIMPLE SAMPLING ALGORITHM

Let $C$ denote the set of ideas and $V$ the participants. Let $m$ and $n$ denote the number of ideas and participants respectively. If participant $i$ prefers idea $x$ to $y$, we denote this by $x \succ_i y$.

---

**ALGORITHM 1:** Approximating the Borda rule

**Input**: $m$ ideas, $n$ participants, a number of samples $N$
**Output**: An output ranking
count$[\cdot] = 0$;
**for** $i \leftarrow 1$ **to** $N$ **do**
  Sample ideas $c_1$ and $c_2$ and a participant $v$ uniformly at random;
  **if** $c_1 \succ_v c_2$ **then**
    count$[c_1]$ = count$[c_1] + 1$;
  **else**
    count$[c_2]$ = count$[c_2] + 1$;

**return** any ranking such that for any $x$ ranked higher than $y$, count$[x] \geq$ count$[y]$;

---

The *Borda score* of an idea $x$ is defined as $s(x) = \sum_{i \in V}(m - r_i(x))$, where $r_i(x)$ denotes the rank that participant $i$ gives to idea $x$. That is, an idea receives $m - 1$ points if it was ranked first in a ranking, $m - 2$ if it was ranked second, and so on. Define the *normalized Borda score* of an idea $x$ to be $n(x) = s(x)/\sum_{x' \in C} s(x')$, so that the sum of all the Borda scores is 1. The *Borda winner* $x^*$ is any idea with the highest score, i.e. $s(x^*) = \max_x s(x)$. Define an $\epsilon$-Borda winner to be any idea $x$ such that $s(x) \geq (1 - \epsilon)s(x^*)$. Define an $\epsilon$-Borda ranking to be any ranking resulting from a normalized score vector $\hat{n}$ such that for any idea $x$, $|\hat{n}(x) - n(x)| \leq 2\epsilon/m$. The Condorcet winner is defined as an idea $x$ which beats all other ideas in a pairwise election. Define an $\epsilon$-Condorcet winner to be an idea $x$ which receives at least $(1 - \epsilon)\frac{n}{2}$ votes against at least $(1 - \epsilon)(m - 1)$ other candidates.

These approximation definitions satisfy a simple interpretation: an $\epsilon$ winner or ranking of a social choice function is one that could have resulted from changing at most $\epsilon$ fraction of the true comparison values involving each idea.

Consider the following sampling algorithm (Algorithm 1). At each step, sample a participant uniformly at random and ask him to compare two ideas sampled uniformly at random. Increment a counter for the idea chosen by the participant and repeat $N$ times. Now form an output ranking by ordering the ideas from those with the highest to lowest counter values, with ties broken arbitrarily.

THEOREM 1.1. *For any $\epsilon, \delta \in (0, 1)$, Algorithm 1 with $N = O(\frac{m}{\epsilon^2} \ln \frac{m}{\delta})$ returns an $\epsilon$-Borda ranking with probability at least $1 - \delta$. Also, the top idea in the returned ranking is an $\epsilon$-Borda winner with probability at least $1 - \delta$.*

Though we will not detail it here, a more complex variant of Algorithm 1 is able to find an $\epsilon$-Condorcet winner in $\tilde{O}(\frac{m}{\epsilon^4})$ comparisons with probability $1 - \delta$, where the notation $\tilde{O}(\cdot)$ hides logarithmic factors.

## 2. ALGORITHM 1 IN THE FINNISH EXPERIMENT

In the Finnish experiment, participants were asked to give feedback on ideas submitted for 41 different topics which ranged from having only two submitted ideas to having fifteen. For each topic, participants were given a randomly chosen subset of the ideas to rank with some probability. There were 308 participants, and 72,003 effective comparisons were collected from the ranking data.

We find results slightly stronger than described in Theorem 1.1, summarized in Figure 1. For each topic, we compute $n(x)$ as the total normalized number of collected comparisons for which $x$ won.[1] Algorithm 1 is then simulated by shuffling the collected comparisons randomly so that the ordered comparisons correspond to a sequence of samples of Algorithm 1. We repeat this 100 times and calculate the average value of $\epsilon$ achieved at each point in time. Finally, we find the time at which $\epsilon$ equals $0.05$ and $0.1$ and plot this against the number of ideas in that topic.

---

[1] We note that this is not the true value of $n(x)$ since we did not collect the full set of comparisons. However, it is a good approximation because of the large number of users and the large number of comparisons we sample (over one-eighth of all possible comparisons) at random.

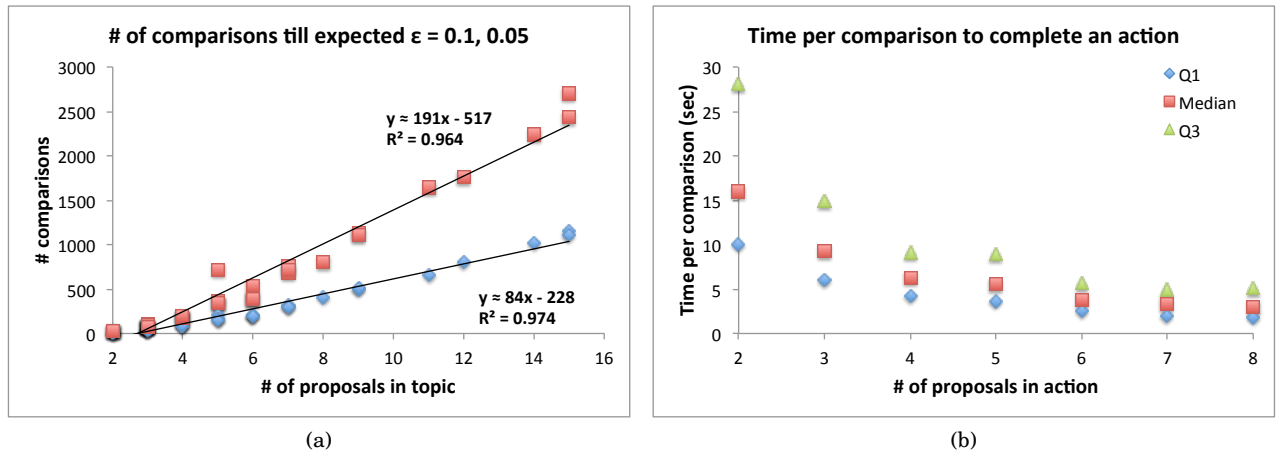

Fig. 1: (a) The red and blue series show the number of comparisons needed to reach an error of $0.05$ and $0.1$ respectively. (b) The time it takes per comparison for participants to complete tasks with varying numbers of proposals.

The main observation to note is that the data series has a good linear fit, and that the constants are reasonable. Given a number of ideas, one can use the linear trend to calculate the total number of comparisons needed to achieve a desired approximation. Since the comparisons are randomly assigned, the expected number of comparisons per participant can be calculated by dividing the resulting number by $n$. For instance, when $x = 100$ and $n = 1000$, the linear trends indicate that $(191 * 100 - 517)/1000 \approx 18.6$ comparisons per participant are needed to achieve $\epsilon = 0.05$ and $(84 * 100 - 228)/1000 \approx 8.2$ are needed for $\epsilon = 0.1$. If one only needs to find winning ideas, one can do even better.

## 3. FURTHER INSIGHTS FROM THE FINNISH EXPERIMENT

Several other insights were found supporting the use of voting rules. We briefly mention them here.

*Bias in rating.* Users were asked to rate some ideas one to five stars. We found that the rating that a user gave to a prior idea greatly influenced the rating given to a subsequent idea. Following a one star rating, a one or five star rating happened 41% and 16% of the time respectively. However, following a five star rating, a one or five star rating happened 22% and 39% of the time respectively. Biases like this may be important factors in evaluating rating-based approaches[Balinski and Laraki 2007].

*Existence of Condorcet winners.* Since a Condorcet winner beats all other proposals in a pairwise comparison, it is typically agreed that when one exists, it should be chosen as the winner. Surprisingly, out of the 41 topics in the Finnish experiment, the vast majority (38) of them had Condorcet winners. This gives several data points indicating that, in practice, one can often expect a Condorcet winner.

*A potential benefit of partial rankings.* In our experiment, we also asked participants to rank small groups of randomly chosen proposals and recorded the amount of time it took participants to complete the action. As shown in Figure 1b, it can be seen that the amount of the time per effective comparison elicited decreases in the range tested, potentially cutting total evaluation time by up to two-thirds.

To conclude, our algorithms and experiments show that social choice functions that were previously thought to place high cognitive burdens on participants can indeed be implemented at scale, a promising sign for the use of crowdsourcing in democratic policy-making.